# Quantifying U-Net Uncertainty in Multi-Parametric MRI-based Glioma Segmentation by Spherical Image Projection

Zhenyu Yang[1,2,3], Kyle Lafata[1,4,5], Eugene Vaios[1], Zongsheng Hu[6,7], Trey Mullikin[1], Fang-Fang Yin[1,2], Chunhao Wang[1]*

[1]Department of Radiation Oncology, Duke University, Durham, NC, USA
[2]Medical Physics Graduate Program, Duke Kunshan University, Kunshan, Jiangsu, China
[3]Medical Physics Graduate Program, Duke University, Durham, NC, USA
[4]Department of Radiology, Duke University, Durham, NC, USA
[5]Department of Electrical and Computer Engineering, Duke University, Durham, NC, USA
[6]Department of Radiation Physics, The University of Texas MD Anderson Cancer Center, Houston, TX, USA
[7]The University of Texas MD Anderson Graduate School of Biomedical Science, Houston, TX, USA

**Short Running Title: UNet Uncertainty by Spherical Projection**

*Corresponding authors:
Chunhao Wang, Ph.D.
Box 3295, Department of Radiation Oncology Duke University Medical Center
Durham, NC, 27710, United States
E-mail: chunhao.wang@duke.edu

**Funding Statement: This work is partially supported by NIH CA014236**
**Conflict of Interest: None associated with this work**

# Abstract

*Background:*

Uncertainty quantification in deep learning is an important research topic. For medical image segmentation, the uncertainty measurements are usually reported as the likelihood that each pixel belongs to the predicted segmentation region. In potential clinical applications, the uncertainty result reflects the algorithm's robustness and supports the confidence and trust of the segmentation result when the ground-truth result is absent. For commonly studied deep learning models, novel methods for quantifying segmentation uncertainty are in demand.

*Purpose:*

To develop a U-Net segmentation uncertainty quantification method based on spherical image projection of multi-parametric MRI (MP-MRI) in glioma segmentation.

*Methods:*

The projection of planar MRI data onto a spherical surface is equivalent to a nonlinear image transformation that retains global anatomical information. By incorporating this image transformation process in our proposed spherical projection-based U-Net (SPU-Net) segmentation model design, multiple independent segmentation predictions can be obtained from a single MRI. The final segmentation is the average of all available results, and the variation can be visualized as a pixel-wise uncertainty map. An uncertainty score was introduced to evaluate and compare the performance of uncertainty measurements.

The proposed SPU-Net model was implemented on the basis of 369 glioma patients with MP-MRI scans (T1, T1-Ce, T2, and FLAIR). Three SPU-Net models were trained to segment enhancing tumor (ET), tumor core (TC), and whole tumor (WT), respectively. The SPU-Net model was compared with (1) the classic U-Net model with test-time augmentation (TTA) and (2) linear scaling-based U-Net (LSU-Net) segmentation models in terms of both segmentation accuracy (Dice coefficient, sensitivity, specificity, and accuracy) and segmentation uncertainty (uncertainty map and uncertainty score).

*Results:*

The developed SPU-Net model successfully achieved low uncertainty for correct segmentation predictions (e.g., tumor interior or healthy tissue interior) and high uncertainty for incorrect results (e.g., tumor boundaries). This model could allow the identification of missed tumor targets or segmentation errors in U-Net. Quantitatively, the SPU-Net model achieved the highest uncertainty scores for three segmentation targets (ET/TC/WT): 0.826/0.848/0.936, compared to 0.784/0.643/0.872 using the U-Net with TTA and 0.743/0.702/0.876 with the LSU-Net (scaling factor = 2). The SPU-Net also achieved statistically significantly higher Dice coefficients, underscoring the improved segmentation accuracy.

*Conclusion:*

The SPU-Net model offers a powerful tool to quantify glioma segmentation uncertainty while improving segmentation accuracy. The proposed method can be generalized to other medical image-related deep-learning applications for uncertainty evaluation.

# 1. Introduction

Automatic image segmentation is a key research topic in medical imaging analysis[1–3]. Driven by recent developments in algorithms and increased computational power, deep learning has become the major vehicle for improved medical image segmentation[2,4,5]. When translating research and development into real-world clinical applications, the robustness of deep neural network (DNN) predictions must be studied before incorporating it into patient care[6–8]. Classic neural networks are limited by their inability to deliver reliable uncertainty estimation and suffer from over- or under-confidence[7]. Specifically, in medical image segmentation, current DNNs learn from training cases (i.e., paired image data and segmentation ground truths derived from manual delineation) and make segmentation predictions using test cases. However, pixel-wise uncertainty, defined as the likelihood that each pixel belongs in the target segmentation region, is unavailable. Without these uncertainty measurements, predictions may provide a false impression of certainty[9]. Additionally, DNNs are prone to overfitting due to frequently underpowered medical image datasets, thus underscoring the importance of uncertainty estimation for robustness assessment (i.e., model calibration) for deep learning models[6,10,11].

To overcome these issues, researchers are actively working to understand and quantify uncertainty in DNN prediction[12]. In general, solutions fall under two approaches: measuring DNN model-related uncertainty (i.e., epistemic uncertainty) and measuring data-related uncertainty (i.e., aleatoric uncertainty)[13]. A representative technique for model-related uncertainty evaluation uses Bayesian neural networks[12,14] which replace a single weight in the DNN with a probability distribution to produce a probabilistic prediction. The variation in model parameters can be translated to the variation in segmentation predictions, yielding an error margin for each pixel. Monte Carlo dropout[15,16] represents an alternative technique. Here, based on the traditional DNN architecture, the prediction uncertainty is estimated by repeating the segmentation prediction multiple times with a certain number of neurons randomly switched off (i.e., dropped out) with each iteration. Both aforementioned estimation techniques have not been well investigated in medical image segmentation due to their complexity and computational costs. In contrast, data-related uncertainty estimation has been more extensively investigated. Medical image data sources are often inhomogeneous (e.g., high intra- and inter-observer variation, image noise effect, and image acquisition protocol variation)[17] with significant real-world data variability (e.g., unknown/rare data entries[18,19]). These limitations have implications for DNN prediction uncertainties [20]. Test-time augmentation[21,22] is commonly used to determine image segmentation uncertainty by evaluating the different augmentation combinations (e.g., rotation, scaling, flipping, adding noise). However, a standard data augmentation strategy for uncertainty evaluation has yet to be established and this trial-and-error method frequently suffers from poor efficiency.

Given these limitations with previously investigated methodologies, we aimed to develop a novel method to quantify data-related uncertainty of U-Net[23] using multi-parametric MRI (MP-MRI)-based brain glioma segmentation. Our approach was inspired by the image processing techniques used in spherical cameras. The planar MRI images were projected onto a pre-defined spherical surface with multiple projection centers. Our hypothesis was that the variation in segmentation results using these projections with different projection centers could reflect the image-related segmentation uncertainty. We utilized this approach as an equivalence to nonlinear image transformation and combined it with U-Net to quantify the uncertainty of glioma segmentation. The effectiveness of this method was evaluated through the comparison studies presented in this work.

## 2. Materials and Methods

### A. Image Data

The *Brain Tumor Segmentation (BraTS) Challenge 2020* dataset was employed in this work[24]. This dataset includes 369 subjects with either low-grade glioma or glioblastoma. Each subject has four standard MR sequences as an MP-MRI protocol: Fluid Attenuated Inversion Recovery (FLAIR), T1-weighted (T1), contrast-enhanced T1-weighted (T1-Ce), and T2-weighted (T2) sequences. The ground-truth (GT) tumor segmentation, contoured by experienced neurosurgeons, includes three overlapping tumor targets as binarized masks: the enhancing tumor (ET), the tumor core (TC), and the whole tumor (WT). Figure 1 illustrates an example of 4 MR sequences with the corresponding ground-truth segmentations. The *BraTS challenge 2020* dataset also includes the following pre-processing steps: (1) co-registration to the same anatomical template, (2) interpolation to the same resolution $1 \times 1 \times 1$ *mm*, and (3) skull-stripping. To facilitate image projection operations, we also extended the in-plane matrix size of the BraTS image from $240 \times 240$ to $256 \times 256$ by padding zeros around the edges.

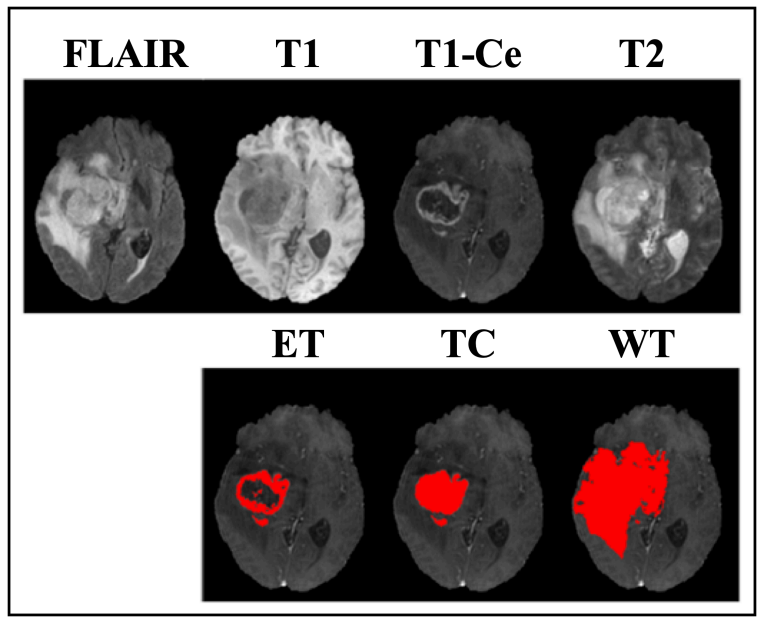

*Figure 1. An example of 4 MR images (i.e., FLAIR, T1, T1-Ce, T2) and the corresponding ground-truth segmentations (i.e., ET, TC, and WT) from the BraTS challenge 2020 dataset.*

## B. Segmentation Model Design

### B.1 Spherical Projection

Inspired by spherical camera image processing[25], we projected the planar images onto a pre-defined spherical surface as part of an image processing step. As illustrated in Figure 2(A), the spherical image projection causes an inhomogeneous scaling over each sub-region of the original image: local image details near the image center are magnified while preserving the field-of-view (FOV) that renders global anatomy[26]. We hypothesize that the spherically projected MR images could be used to quantify segmentation uncertainty. Segmentation variation arising from multiple spherical projections derived from an original image reflects internal uncertainty in the segmentation results.

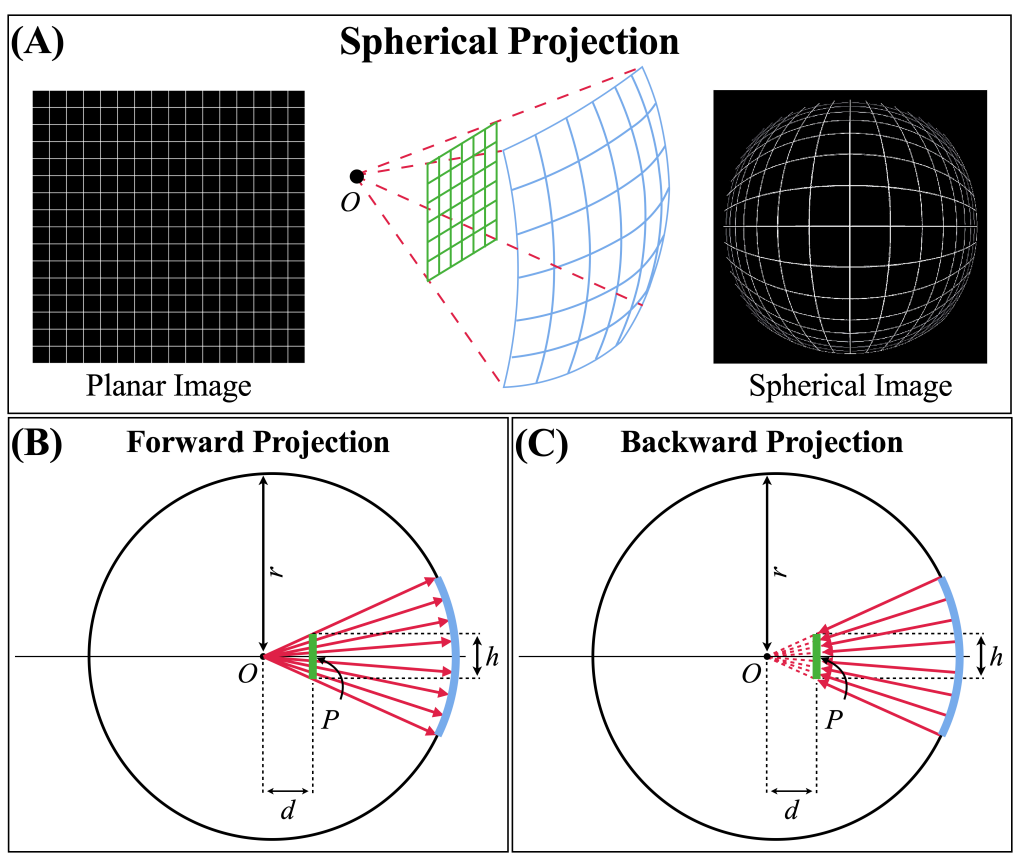

*Figure 2. (A) Schematic diagram of a spherical image projection. (B) Forward projection: projecting a planar image (green) onto a spherical surface as a spherical image (blue). (C) Backward projection: projecting a spherical image (blue) onto a Cartesian plane as a planar image (green).*

Mathematically, we refer to the projection from a planar image to a spherical surface as the forward projection. Assume the sphere in Figure 2(A) has a center $O$ and radius $r$, and $P$ is the center of the planar image. The lateral view of the forward projection is shown in Figure 2(B). The planar image is placed perpendicular to line $OP$ at projection distance $d$. The spherical image can be obtained by projecting each pixel within the planar image onto the pre-defined sphere along its radius. The lateral view of the backward projection, i.e., the inverse operation of the forward projection, is shown in Figure 2(C). Given a planar image with a size of $h \times h$, the relationship of $h$, $d$, and $r$ jointly determines the projection geometry. Without loss of generality, we let the sphere be unit $r = 1$, and $h = 0.5$. As such, the projection geometry, for both forward projection and backward projection, is governed by the projection distance $d$.

To quantify the locoregional scaling effect of the spherical projection, a $16 \times 16$ lattice, as shown in Figure 2(A), was created with image dimensions matching the *BraTS* dataset (i.e., $256 \times 256$ matrix size). This lattice image $I$ was first forward-projected as $I'$ and then backward-projected as $I''$. The locoregional scaling factor $f_{SP}$ was defined as the area ratio of each square tile within $I'$:

$$f_{SP} = \frac{\text{area of each square aure tile in } I'}{\text{area of each square tile in } I \ (= 16 \times 16)} \tag{1}$$

Implementing spherical projection on images with a finite resolution requires interpolation calculations; consequently, the paired forward and backward projection operations degrade the quality of $I''$ (in reference to $I$). In this work, we employed the structural similarity index (SSIM) to determine the optimal projection distance setting: the SSIM was calculated within the central $8 \times 8$ lattice region of $I$ and $I''$, and projection distance $d$ with the highest SSIM value (i.e., minimal image quality degradation) was adopted in the following MP-MRI processing.

*B.2 Segmentation Model Design*

The overall design of the spherical projection-based U-Net (SPU-Net) model is summarized in Figure 3. A U-Net DNN, with its encoding and decoding parts as shown in Figure 3(B), was first constructed, and it has been widely implemented for image segmentation tasks[27,28]. The encoding part consists of repeated convolutional layers, each followed by a max pooling operation. This hierarchical operation encodes the image into multiple levels of feature representations. The decoding part includes repeated up-convolutional layers. Obtained deep image features are then concatenated with the encoding part. This design projects discriminative features onto the image space to obtain a pixel-wise classification. Another $1 \times 1$ convolutional layer with a sigmoid operation is followed to produce a segmentation prediction as a probability distribution: the pixel value is the probability of belonging to the correct segmentation in the range of 0 to 1. Because spherically projected images were processed as nonlinear image transforms within a Cartesian grid, we coded the SPU-Net architecture within the same grid to streamline its implementation.

We denote the original input 4-channel MP-MRI as $X$ (dimension = $256 \times 256 \times 4$), with the center shown as a green dot. Given a projection distance $d$, a single forward projection causes an inhomogeneous scaling to each sub-region. To achieve a uniform transformation effect across $X$, multiple forward projections can be performed with different projection centers. As shown by the white dots in Figure 3(A), the projection centers are set to be evenly distributed across $X$, while zero padding is adopted to maintain the FOV size during the forward projection. Let $n \in \mathbf{N} = \{1, 2, 3, ..., K\}$, where $K$ is the total number of available projection centers. The SPU-Net model for a given MR image $X$ can be summarized as follows:

1) By performing the forward projection with $k$ different image centers, a set of $k$ 4-channel forward-projected images $\mathbf{X}_S = \{X_S^n\}_{n \in \mathbf{N}}$ can be obtained as a $k \times 256 \times 256 \times 4$ tensor.

2) A set of $k$ probability distribution maps $\mathbf{M}_S = \{M_S^n\}_{n \in \mathbf{N}}$ are generated as the output by U-Net using $\mathbf{X}_S$ on the spherical surface.

3) After backward projection, A set of $k$ probability distribution maps $\mathbf{M} = \{M^n\}_{n \in \mathbf{N}}$ are obtained from $\mathbf{M}_S$ on the Cartesian plane.

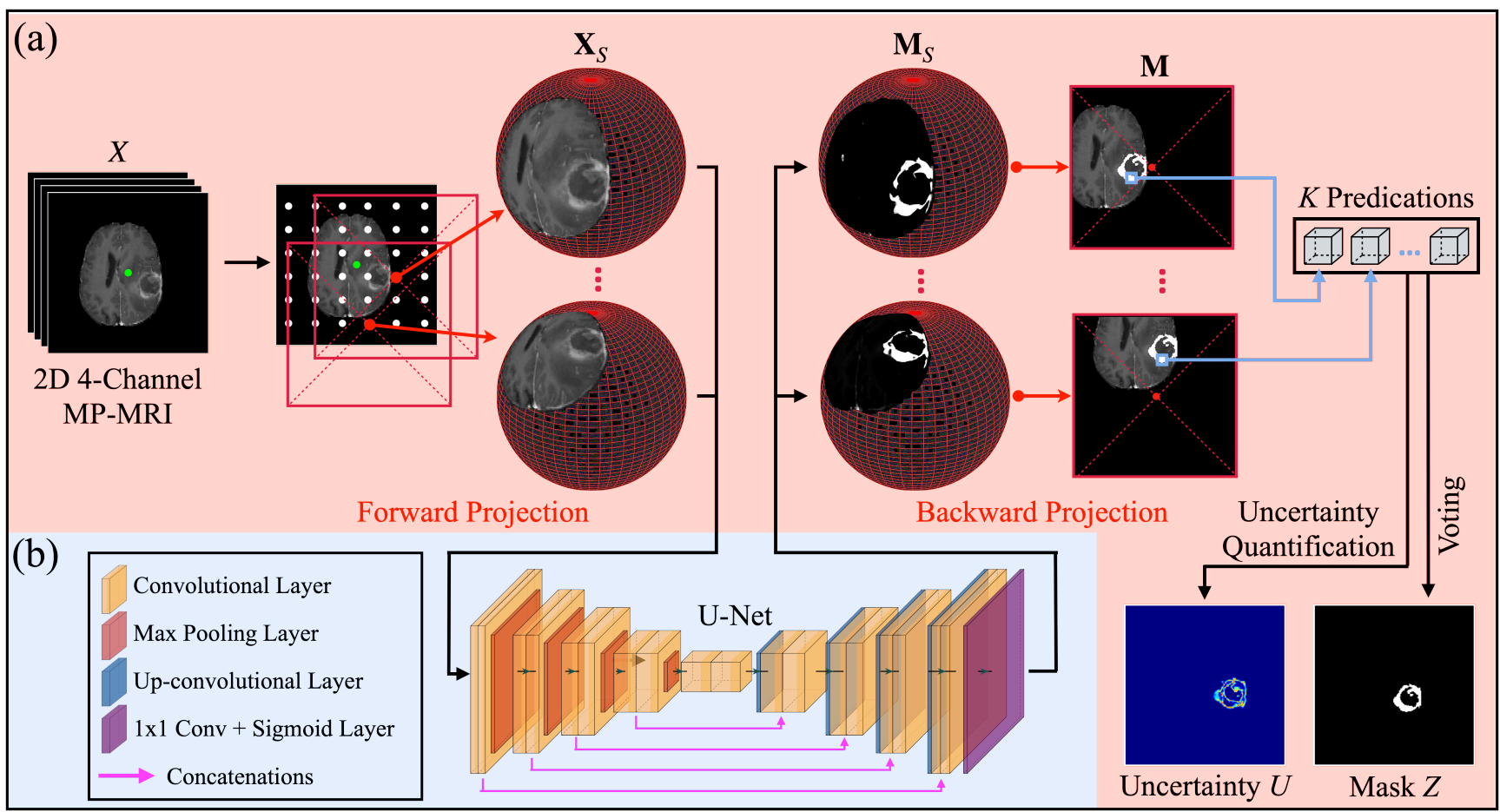

*Figure 3. (A) The overall design of the proposed spherical projection-based U-Net (SPU-Net) model. Given a 2D MRI image X, the forward projection is performed to generate a set of spherical images $\mathbf{X}_S$, and the corresponding segmentation $\mathbf{M}_S$ can be obtained from $\mathbf{X}_S$ by U-Net. The backward projection is subsequently performed to project $\mathbf{M}_S$ onto the Cartesian plane as $\mathbf{M}$. Finally, the segmentation uncertainty U and binarized segmentation mask Z can be obtained by analyzing $\mathbf{M}$. (B) U-Net architecture.*

For each pixel $(i, j)$ within $X$, $K$ independent segmentation predictions can be found within $\mathbf{M}$, i.e., $\mathbf{M}_{(i,j)} = \{M_{(i,j)}^1, M_{(i,j)}^2, M_{(i,j)}^3, \dots, M_{(i,j)}^K\}$. To reach the final segmentation result as a binarized mask $Z$, a voting scheme is designed:

$$\begin{cases} \text{if } \left(\sum_{n=1}^{K} M_{(i,j)}^n \right)/K > 0.5, & Z_{(i,j)} = 1 \\ \text{otherwise}, & Z_{(i,j)} = 0 \end{cases}, \qquad (2)$$

where $Z_{(i,j)}$ is the value of pixel $(i, j)$ in $Z$. The uncertainty $U$ can be quantified by the entropy of the segmentation predictions[10]. Suppose there are $T$ unique values in $\mathbf{M}_{(i,j)}$, and the frequency of the $t$-th unique value is $\hat{p}_{(i,j)}^t$. As such, the uncertainty can be approximated as:

$$U_{(i,j)} = -\sum_{t=1}^{T} \hat{p}_{(i,j)}^t \ln\left(\hat{p}_{(i,j)}^t\right) \qquad (3)$$

where $U_{(i,j)}$ is the uncertainty at pixel $(i, j)$ in $U$. Note that $\mathbf{M}_{(i,j)}$ contains continuous numbers between 0 and 1. Here, we normalize and discretize $\mathbf{M}_{(i,j)}$ to the range of 0-100 for the ease of entropy computation[10,29].

Based on the pixel-wise segmentation uncertainty result, a quantitative evaluation metric is needed to evaluate the overall image segmentation uncertainty. This metric is expected to be high when (1) $U_{(i,j)}$ is low in the pixel where the segmentation result $Z_{(i,j)}$ is correct (i.e., high confidence in correct results), and (2) $U_{(i,j)}$ is high in the pixel where $Z_{(i,j)}$ is incorrect (i.e., reasonable doubt in wrong results). In this work, we adopted the uncertainty score defined in the *BraTS segmentation uncertainty challenge 2020* (*QU-BraTS challenge 2020*)[29]. Specifically, the obtained $U$ is first linearly normalized to 0-100, and an uncertainty threshold $\tau$ is set to 100 predetermined integers as $\tau = 1, 2, \dots, 100$. At each threshold $\tau$, all pixels with uncertainty value $U_{(i,j)} \geq \tau$ are marked as "uncertain", and the associated segmentation results $Z_{(i,j)}$ are filtered out and not considered in the subsequent calculations. The remaining segmentation results in $Z_{(i,j)}$ are compared with the corresponding ground truth based on the Dice similarity index. At each $\tau$, the total number of true positive pixels and true negative pixels can also be obtained in the remaining $Z_{(i,j)}$ as $\text{TP}_\tau$ and $\text{TN}_\tau$, respectively. The ratio of filtered true positive (FTP) pixels at threshold $\tau$ is defined as $FTP = (TP_{100} - TP_\tau) / TP_{100}$, where the $TP_{100}$ is the number of true positive pixels in the unfiltered $Z$ (i.e., $\tau = 100$). The ratio of filtered true negative (FTN) pixels is defined in a similar manner. As such, three curves and their corresponding area under the curve (AUC) can be obtained:

1) $AUC_1$: the area under the curve of the Dice similarity index versus $\tau$. High $AUC_1$ indicates that segmentation predictions are accurate in regions of low uncertainty (i.e., high confidence for the correct results).

2) $AUC_2$: the area under the curve of FTP versus $\tau$. $(1 - AUC_2)$ penalizes the low confidence in the true positive predictions.

3) $AUC_3$: the area under the curve of FTN versus $\tau$. $(1 - AUC_3)$ penalizes the low confidence in the true negative predictions.

Collectively, the final uncertainty score is defined as:

$$\text{Score} = \frac{AUC_1 + (1 - AUC_2) + (1 - AUC_3)}{3}. \tag{4}$$

The uncertainty score (1) rewards high confidence in correct predictions and low confidence in incorrect predictions, and (2) penalizes low confidence for pixels with correct predictions[29].

## C. Comparison Study

In the adopted *BraTS* 2020 dataset with 369 subjects, axial 2D images from the four MR sequences were used as 4-channel 2D samples. The sample usage for training and independent test follows the 8:2 ratio in the patient assignment, and five-fold cross-validation within the training set was employed. In the proposed SPU-Net model shown in Figure 3, the projection geometry was determined by the optimal $d$ selection in the study of B.1, and the projection center in Figure 3(A) was spaced at 8-pixel intervals (Hence $K = 1024$). During the training, the loss function was binary cross-entropy, and the Adam optimizer with an initial learning rate of $10^{-3}$ was adopted. The segmentation accuracy was evaluated by the sensitivity, specificity, accuracy, and Dice similarity index. The segmentation uncertainty was evaluated by the uncertainty score in Equation (3). While a single model can be used to simultaneously segment ET, TC, and WT, the inter-correlation between the three targets may limit uncertainty quantification. Therefore, we developed three independent SPU-Net models for ET, TC, and WT segmentation, respectively. During the training, the loss function was binary cross-entropy, and the Adam optimizer with an initial learning rate of $10^{-3}$ was adopted. The segmentation accuracy was evaluated by the sensitivity, specificity, accuracy, and Dice similarity index. The segmentation uncertainty was evaluated by the uncertainty score in Equation (4).

In the comparison study, two additional segmentation models were studied:

1) Classic U-Net model with test-time augmentation (TTA). Specifically, the TTA protocol includes image rotation, flipping, scaling, and noise addition. The augmentation is repeated $k = 1024$ times to match our SPU-Net model.

2) Linear scaling-based U-Net (LSU-Net) model. Since our spherical projection operation is equivalent to an inhomogeneous image scaling, it is worth comparing the homogeneous linear upscaling effect with our spherical projection design. The model design was summarized in Figure 4, the linear upscaling was employed to magnify the locoregional image content at $k$ different image centers. Cropping was subsequently employed to maintain the input tensor dimension as $k \times 256 \times 256 \times 4$. The rest of the design followed the SPU-Net model, and the final segmentation result with uncertainty was evaluated following Equations (2)-(3). The linear scaling factor $f_{LS}$ was set close to $f_{SP}$ near the image center in the spherical projection design (as Figure 2(A)).

In these two models, the training settings, including training/test set assignment, loss function, and initial learning rate, were kept the same as the proposed SPU-Net model. The achieved segmentation accuracy (sensitivity, specificity, accuracy, and Dice similarity index) and uncertainty score (in Equation (4)) were compared by the Wilcoxon signed-rank test. The statistical significance level was set at 0.05.

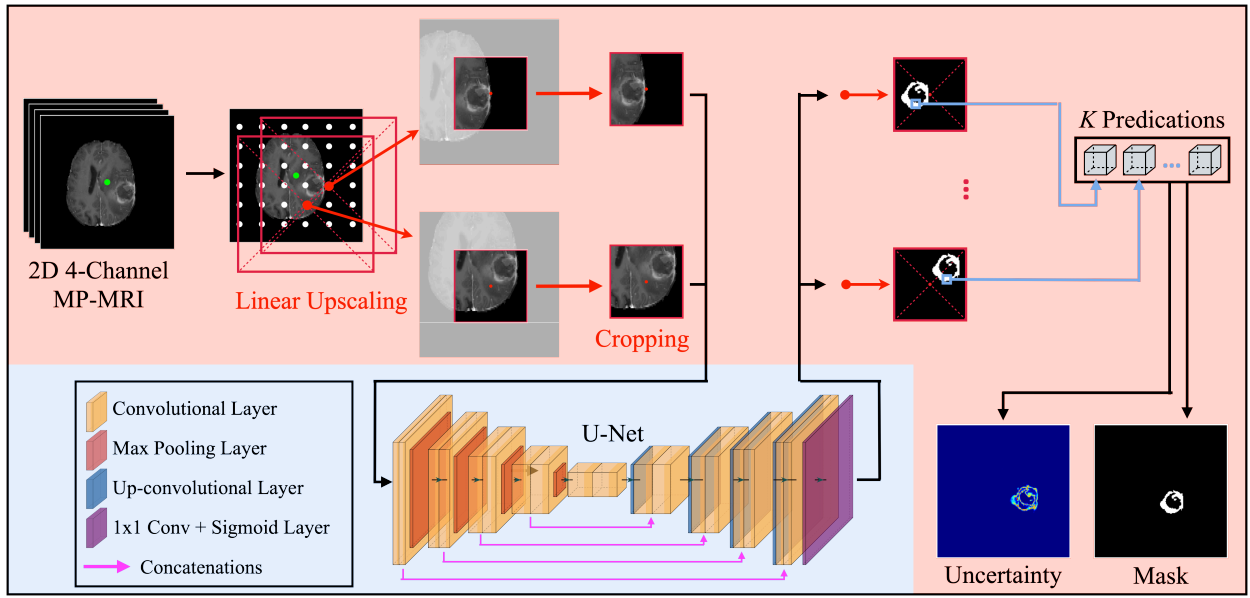

*Figure 4. The linear scaling-based U-Net (LSU-Net) model for comparison purposes. The workflow follows SPU-Net shown in Figure 3.*

## 3. Results

Figure 5(A) shows the SSIM as a function of projection distance $d$. As shown, the paired forward and backward projections at $d = 0.3$ produce minimal image quality degradation near the image center, and thus $d = 0.3$ was adopted in our design. Figure 5(B) illustrates the created $16 \times 16$ lattice image $I$, its forward-projected image $I'$, and the corresponding backward-projected image $I''$ at $d = 0.3$. Our in-house forward projection algorithm upscaled the lattice size near the image center while preserving the whole lattice structure. The backward projection successfully restored the overall lattice integrity from $I'$. The locoregional scaling effect of the spherical projection at $d = 0.3$ was quantified in Figure 5(C), i.e., the area ratio of each square tile in $I$ and $I'$. The factor $f_{SP}$ near the image center ranged from 2 to 3 (as marked by the red box); therefore, we studied linear upscaling factors $f_{LS} = 2$ and $f_{LS} = 3$ in the LSU-Net model for comparison purposes.

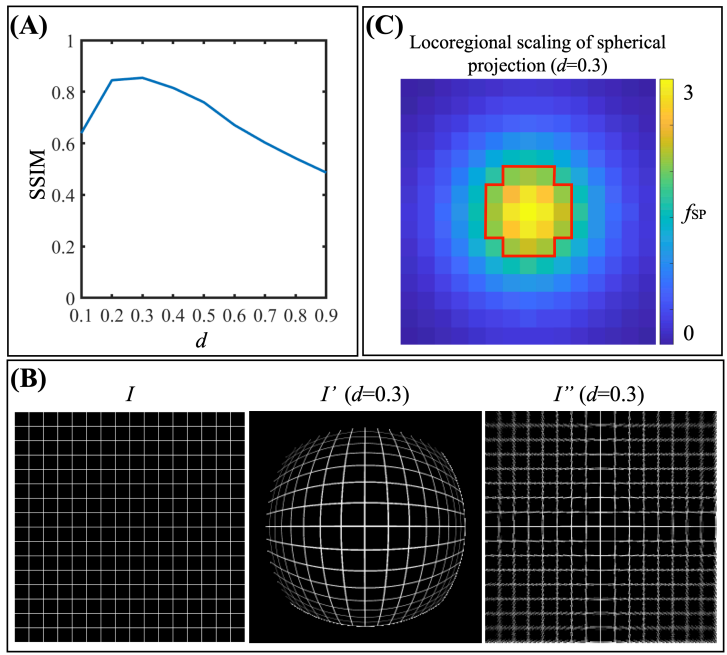

*Figure 5. (A) SSIM (between the central $8 \times 8$ lattices of I and I'') as a function of the projection distance $d$. (B) The created $16 \times 16$ lattice image I, its forward-projected image I', and the corresponding backward-projected image I'' at $d = 0.3$. (C) The locoregional scaling effect of the spherical projection at $d = 0.3$ (i.e., the area ratio of each square tile in I and I' at $d = 0.3$.)*

Figure 6 shows an example of ET segmentation from the classic U-Net model, the LSU-Net model, and the proposed SPU-Net model. The original MRIs with ground truth segmentation are shown in the left panel. The binarized segmentation masks with pixel-wise segmentation uncertainty are rendered in the right panel. The segmentation mask from the SPU-Net model demonstrates the highest visual consistency with the ground truth, and the classic U-Net model shows good visual consistency with some discrepancies near the tumor rim. For both $f_{LS} = 2$ and $f_{LS} = 3$, the LSU-Net models show low sensitivity in the tumor region and falsely identified the contrast-enhanced blood vessels on the right side as tumors. Among the four models, the LSU-Net ($f_{LS} = 3$) model shows the largest visual disagreement with the ground truth. The SPU-Net model's uncertainty is (1) low in the tumor interior and normal tissue interior, and (2) high on the segmentation mask's boundary. The missing tumor region in the mask (as marked by the red arrows) is appropriately indicated in the uncertainty map $U$. These results highlight the potential segmentation variation that follows human operations perception, which is consistent with our expectations: minimal uncertainty in the correct segmentation region and high uncertainty in the incorrect segmentations. In contrast, the uncertainty from the LSU-Net model (for both $f_{LS} = 2$ and $f_{LS} = 3$) fails to render a definite pattern. Although blurred contours can be observed, the numerical values are insufficient to highlight the boundary. The segmentation uncertainty from the classic U-Net model is better than the LSU-Net model results, but several correct segmentation predictions (e.g., tumor interior) are marked with high uncertainty. Figures 7 and 8 illustrate the TC and WT segmentations, respectively. The superior SPU-Net results in Figure 6 are similarly appreciated.

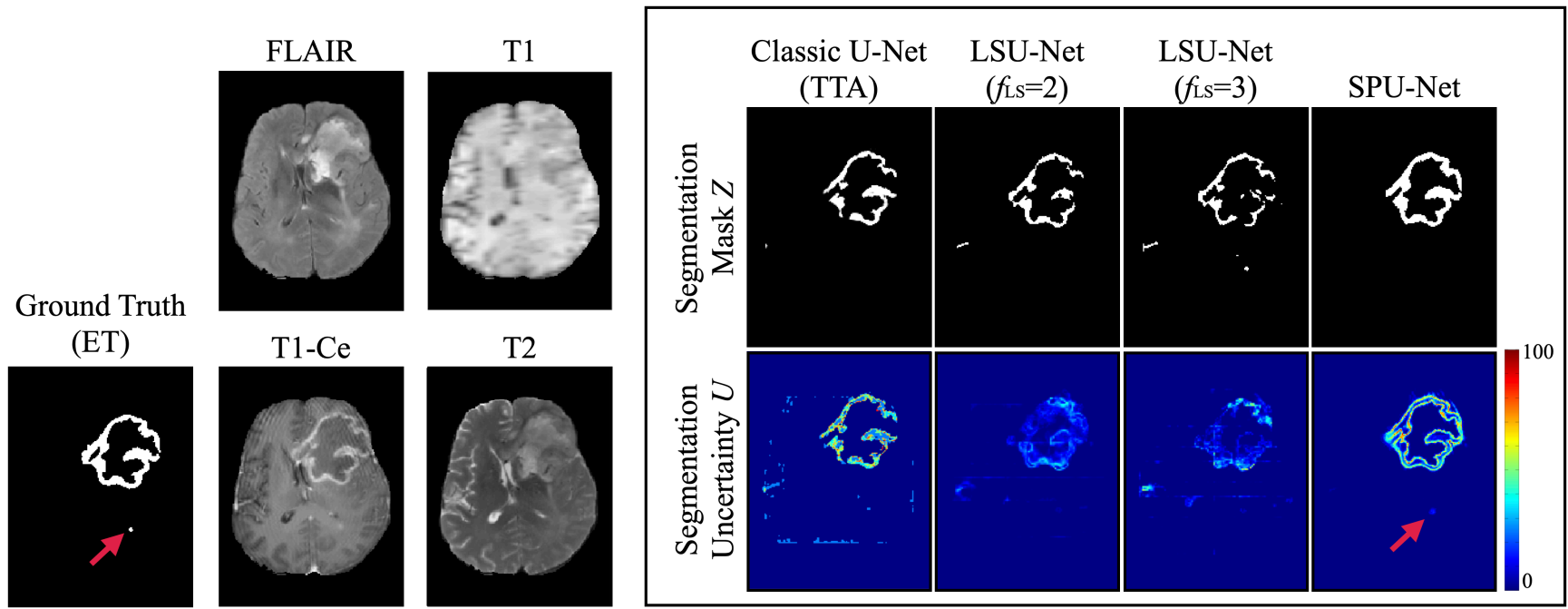

Figure 6. *An example of ET segmentation from the classic U-Net model, the LSU-Net ($f_{LS}$ = 2) model, the LSU-Net ($f_{LS}$ = 3) model, and the SPU-Net model. The binarized segmentation mask and pixel-wise segmentation uncertainty are demonstrated for each model.*

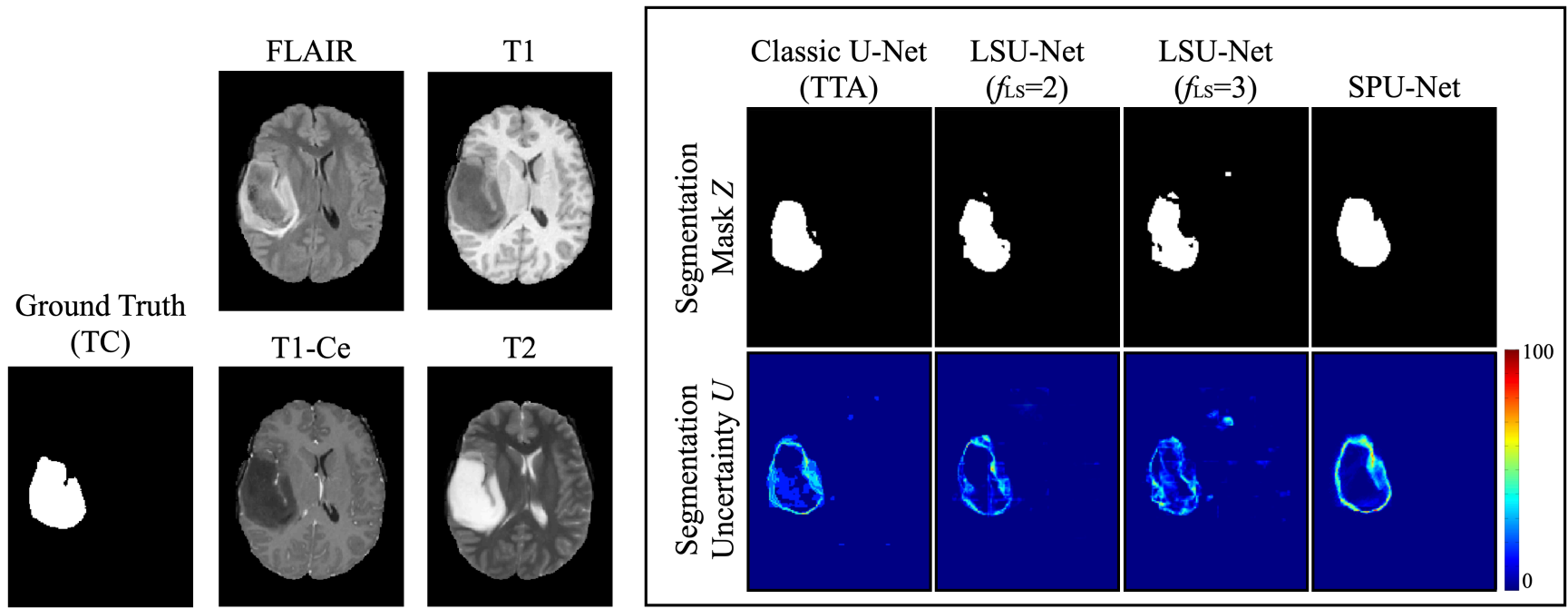

*Figure 7. An example of TC segmentation (binarized segmentation results and the corresponding segmentation uncertainty).*

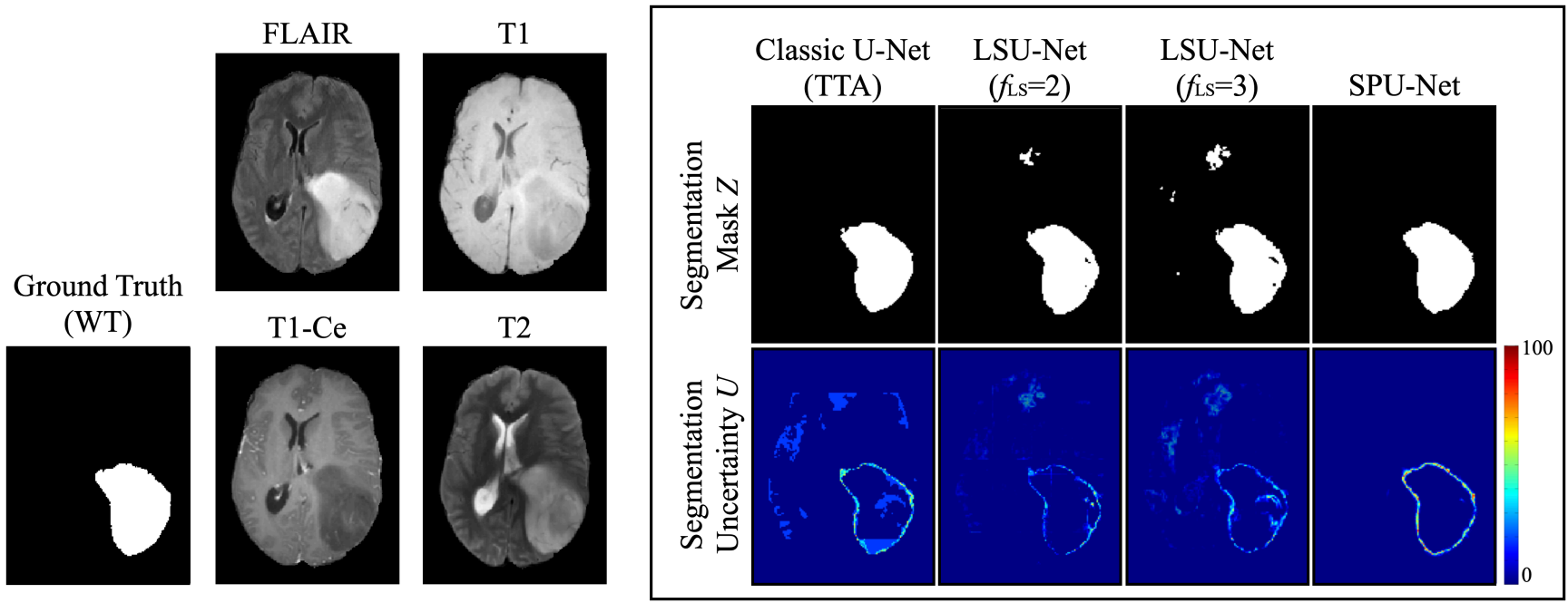

*Figure 8. An example of WT segmentation (binarized segmentation results and the corresponding segmentation uncertainty).*

Figure 9(A)-(C) shows the curves of the Dice similarity index, FTP, and FTN as a function of $100 - \tau$ for three segmentation targets (i.e., ET, TC, and WT), respectively. The blue, yellow, red, and green curves correspond to the classic U-Net, LSU-Net ($f_{LS} = 2$), LSU-Net ($f_{LS} = 3$), and SPU-Net models, respectively. Each curve is represented as a shaded plot, where the solid line represents the mean value for all test cases and the shaded area indicates the standard deviation. As illustrated, the SPU-Net model has a higher Dice curve for all three segmentation targets compared to the rest models, producing the highest mean $AUC_1$. Dice curves of the LSU-Net ($f_{LS} = 2$) model are close to the classic U-Net model for three segmentation targets, while LSU-Net ($f_{LS} = 3$) yields the lowest results. Note that the Dice index of the SPU-Net model is very high even when $\tau$ is very small. This finding suggests that the SPU-Net model produces an accurate segmentation prediction with very high confidence (i.e., very low uncertainty). As $\tau$ increases (i.e., involving more "uncertain" pixels into consideration), the Dice curve has a decreasing slope for ET and TC segmentation and a steady slope for WT segmentation. These results suggest that segmentation prediction can be less accurate for those "uncertain" pixels, which is consistent with our expectations. For the FTP results, the SPU-Net model achieves the lowest mean $AUC_2$, which indicates that true positive predictions are obtained with high confidence. Based on the observation of the FTP curve, all models show a similar shape with different $AUC_2$ results. As $\tau$ approaches 100, the least true positive predictions are filtered out in our SPU-Net model for all segmentation targets. For the FTN results, all models achieved a very steady slope with a low mean $AUC_3$, indicating that true negative predictions are determined with high confidence.

Table I summarizes the segmentation accuracy evaluations and uncertainty scores from five-fold cross-validation (mean ± standard deviation) from the four models. In terms of segmentation accuracy, the proposed SPU-Net model achieved the highest mean Dice similarity index for all three targets. In contrast, the LSU-Net model did not show a robust improvement compared to the classic U-Net model. With respect to uncertainty scores, the proposed SPU-Net model achieved a significantly improved score. Though the LSU-Net ($f_{LS} = 2$) model achieved higher scores across all three segmentations compared to LSU-Net ($f_{LS} = 3$), LSU-Net did not consistently outperform the classic U-Net model.

*Table I. Five-fold cross-validation segmentation results and uncertainty score (mean ± standard deviation) for three segmentation targets from the classic U-Net model, the LSU-Net model, and the SPU-Net model.*

|  |  | **Accuracy** | **Sensitivity** | **Specificity** | **Dice** | **Uncertainty Score** |
|---|---|---|---|---|---|---|
| **ET** | U-Net (TTA) | 0.9757±0.1393 | 0.7977±0.1898 | 0.9785±0.1402 | 0.8031±0.1974* | 0.7837±0.1546* |
|  | LSU-Net ($f_{LS}$=2) | 0.9923±0.0512 | 0.7952±0.1654 | 0.9952±0.0514 | 0.8104±0.1667* | 0.7431±0.1586* |
|  | LSU-Net ($f_{LS}$=3) | 0.9906±0.0561 | 0.7554±0.1664 | 0.9944±0.0563 | 0.7807±0.1734* | 0.7170±0.1488* |
|  | SPU-Net | 0.9874±0.0988 | 0.8862±0.1347 | 0.9887±0.0997 | **0.8820±0.1478** | **0.8262±0.1643** |
| **TC** | U-Net (TTA) | 0.8542±0.3241 | 0.7318±0.3081 | 0.8613±0.3419 | 0.6433±0.3380* | 0.6428±0.2614* |
|  | LSU-Net ($f_{LS}$=2) | 0.9422±0.1856 | 0.7027±0.2922 | 0.9546±0.1952 | 0.7068±0.2908* | 0.7015±0.2302* |
|  | LSU-Net ($f_{LS}$=3) | 0.9384±0.1833 | 0.6521±0.2949 | 0.9515±0.1973 | 0.6455±0.2899* | 0.6565±0.2174* |
|  | SPU-Net | 0.9458±0.1975 | 0.8168±0.2280 | 0.9530±0.2036 | **0.7966±0.2514** | **0.8479±0.1487** |
| **WT** | U-Net (TTA) | 0.9892±0.0244 | 0.8838±0.1202 | 0.9958±0.0237 | 0.8821±0.0889* | 0.8715±0.0962* |
|  | LSU-Net ($f_{LS}$=2) | 0.9871±0.0111 | 0.8703±0.0995 | 0.9944±0.0069 | 0.8858±0.0849* | 0.8757±0.0637* |
|  | LSU-Net ($f_{LS}$=3) | 0.9811±0.0166 | 0.8268±0.1136 | 0.9910±0.0120 | 0.8413±0.1039* | 0.8362±0.0672* |
|  | SPU-Net | 0.9928±0.0175 | 0.9303±0.0746 | 0.9905±0.0169 | **0.9365±0.0603** | **0.9359±0.0495** |

* statistically significant result compared to the SPU-Net model

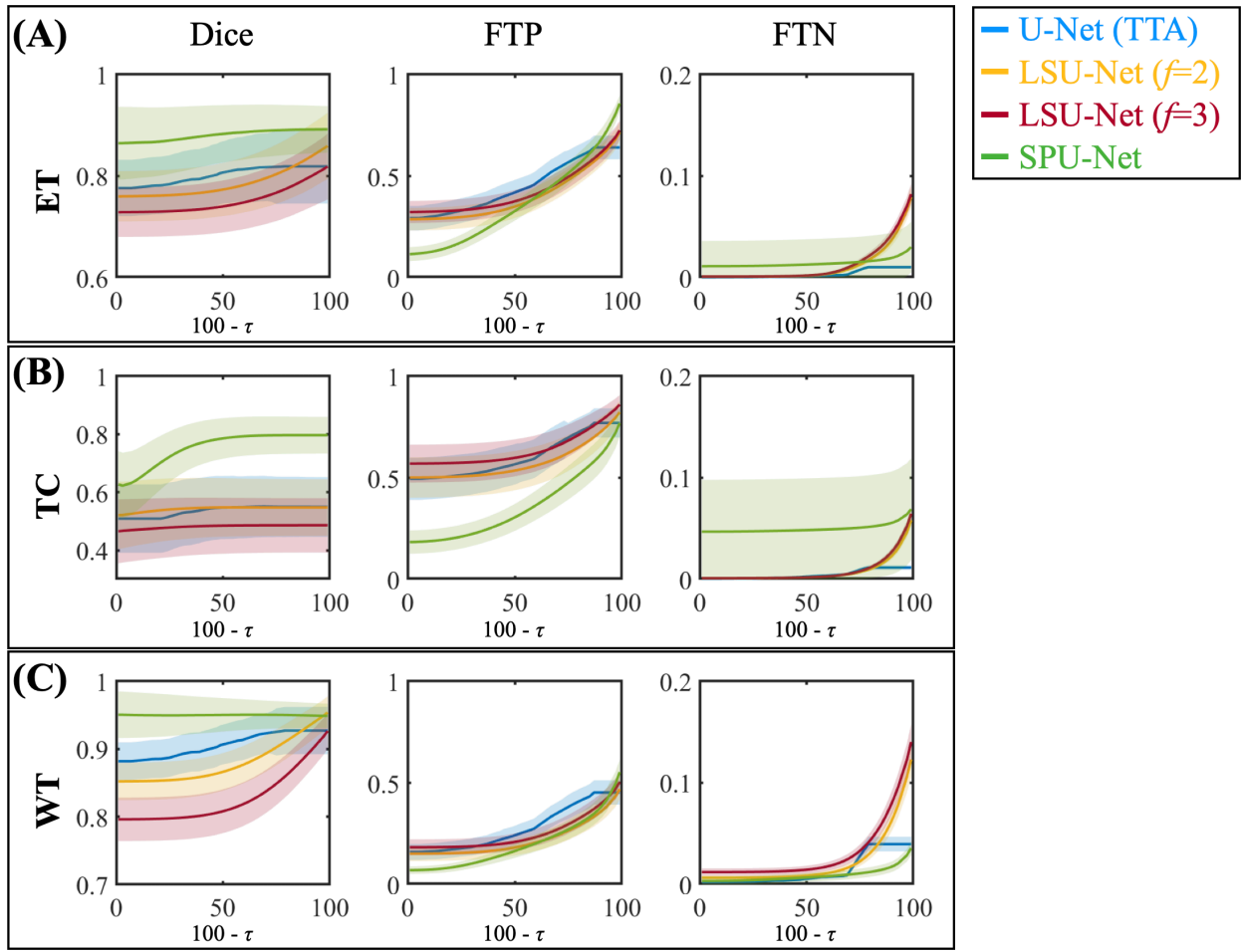

*Figure 9. (A)-(C) The curve of the Dice similarity index, FTP, and FTN as a function of 100-τ for ET, TC, and WT segmentation, respectively.*

## 4. Discussion

We developed a novel U-Net segmentation uncertainty quantification method using spherical projection for glioma segmentation using MP-MRI sequences obtained from a large cohort of patients in the BraTS database. A key innovation in this work is the ability to project planer MR images onto a spherical surface as part of a proposed U-Net segmentation workflow, which is equivalent to a nonlinear image transformation. In a group of projections with different centers across the entire FOV, fine structures are magnified with varying scales, resulting in non-linear transform effects. Multiple independent segmentation predictions can then be obtained by U-Net from a single MR image. Our hypothesis is that the segmentation prediction resulting from spherical transformation may mirror the uncertainty present in the model, such that a high degree of consistency in the image predictions across multiple transformations corresponds to a high level of confidence. Pixel-wise segmentation uncertainty can be obtained and visualized as an anatomical image. The uncertainty estimation improves the interpretability of a binarized segmentation mask. Figure 6 depicts how this model accurately highlights missed tumor regions and potential segmentation errors with high uncertainty. In addition, our design is different from the attention mechanisms, which refer to special model designs that guide a model to concentrate on relevant portions of the input data[30]. Most research integrating deep learning with visual attention mechanisms in contemporary CNN models employs masking techniques to pinpoint critical channel-wise or spatial-wise image features via another layer, i.e., attention modules, which have trainable weights[31]. Our spherical projection technique focuses on manipulating input image data, and it thus becomes different from the attention mechanism.

Image Segmentation accuracy and uncertainty quantification are two separate yet related topics: for an image segmentation model, the segmentation accuracy is the absolute difference between the model's output and the ground-truth results when evaluating the model with an independent test set; on the other hand, the segmentation uncertainty aims to quantify the confidence of the model when generating output. Previous studies demonstrated that high uncertainty regions near tumor boundaries reflect intra- and inter-observer variation that occurs with manual delineation[32]. Figure 9 shows that the correct predictions (for both true positive and true negative results) can be obtained with high degrees of confidence. These findings, along with the uncertainty score, suggest that the SPU-Net model outperforms the classic U-Net and LSU-Net models with regard to uncertainty quantification. In the proposed SPU-Net model, each sub-region within the original MRI is transformed into a series of projected images with varied magnifications. The U-Net model predicts the segmentation (for a given sub-region) under diverse appearances, which may better reflect internal ambiguity with segmentation results. In contrast, the TTA in the classic U-Net model applies the global transformation to the entire image without variations of locoregional image details. These uncertainty results were inferior to results from the proposed SPU-Net model. While the linear upscaling design in the LSU-Net models may enhance the locoregional image details, it ignores the unscaled peripheral image features; consequently, the locoregional transformation, even after repetition for full FOV coverage, is insufficient for uncertainty quantification.

The *BraTS challenge 2020*[29,33] has been extensively studied, with numerous deep learning architectures reporting state-of-the-art segmentation accuracy. Despite its use of the classical U-Net structure with a smaller number of parameters, our proposed SPU-Net model achieved an improved or comparable Dice coefficient, demonstrating the effectiveness of our spherical projection technique. In terms of segmentation uncertainty, the *QU-BraTS challenge 2020*[29] benchmarked 15 different uncertainty quantification methods, producing uncertainty scores ranging from 0.5828-0.8885, 0.5989-0.9135, and 0.6312-0.9429 for ET, TC, and WT segmentation, respectively. Compared to the benchmarked models, our SPU-Net achieved comparable uncertainty scores to the top 5 models of each segmentation task. Notably, the SPU-Net results and the aforementioned state-of-the-art results come from different independent test sets. Thus, a direct comparison is unfortunately unavailable. Furthermore, our spherical projection technique can be easily integrated into other deep-learning segmentation models. As shown in the *Supplementary Materials*, two other models with the proposed spherical projection design were investigated, namely SPU-Net++ (i.e., U-Net++[34] with spherical projection images) and SP-MH-UNet (i.e., MH-UNet[35] with spherical projection images). We observed that both models with the proposed spherical projection design exhibited substantial increases in the uncertainty score, suggesting enhanced uncertainty quantification in all three segmentation tasks. Additionally, both models achieved improved or similar segmentation accuracies (measured by the Dice coefficient) compared to their original versions. These results suggest that our proposed spherical projection method can serve as a model-independent technique for quantifying segmentation uncertainty, while also potentially improving segmentation accuracy.

In general, DNN models fundamentally rely on sufficient, homogeneous, annotated data[36]. Due to limitations with currently available image datasets, the development of an accurate segmentation tool for clinical application faces significant hurdles [37]. Better triaging of simple and complex cases between humans and computer algorithms could bridge gaps in current DNN models[38]: DNN models could be leveraged to analyze simple cases with high confidence and reliability, while difficult ones with high observable uncertainty should be referred to experienced radiologists. In this study, our SPU-Net model was approved to be a powerful tool to guide the clinical review of segmentations with high ambiguity. By replacing the U-Net with other DNN structures (such as Bayesian neural networks or Monte Carlo dropout), our SPU-Net model can be incorporated with other existing uncertainty quantification methods for uncertainty characterization. Furthermore, the proposed spherical projection-based image transformation can be generalized to other medical image-related deep learning applications for uncertainty evaluation (e.g., classification, regression, etc.).

In addition to the uncertainty quantification, the SPU-Net model improves glioma segmentation accuracy. Similar to data augmentation strategies in deep learning, the spherical projection provides a large data sample size with a diverse appearance, which can be used to improve model robustness and generalizability. Previous studies reported that small tumors/organs with fine structures and complex boundaries are challenging for the classic U-Net[39]. This is evident in our results in Figure 6. One explanation of such limitation is that the U-Net adopts the fixed receptive field of the convolution kernel[40]. Typically, a large receptive field ignores the small structure, whereas a small receptive field may extract redundant image features. Gliomas can manifest in a wide variety of sizes, shapes, and locations across the brain[41], and a single convolutional kernel thus may be insufficient for every instance[42,43]. Previous research has shown that combining the convolutional kernels with different sizes benefits segmentation. Our proposed spherical projection model addresses this challenge by magnifying the locoregional image content at different scales, effectively altering the receptive fields relative to the entire brain without modifying the deep neural network structure. By combining the results from multiple receptive fields, the final segmentation mask is more accurate, even for complex tumor boundaries, as illustrated in Figure 6. Another interesting finding is that the LSU-Net ($f_{LS} = 3$) model consistently underperforms compared to the LSU-Net ($f_{LS} = 2$) model. As $f_{LS}$ increases, more peripheral image content outside the upscaled region is lost; in contrast, the spherical projection preserves the global anatomical information within FOV (as in Figure 2(A)), which improves segmentation accuracy.

As a feasibility study, the current spherical projection was designed and implemented using a pre-defined spherical surface. Given a projection distance $d$, the image after the forward or backward projection was determined. The optimal $d$ was subsequently investigated by measuring the degradation in image quality following a paired forward and backward projection. Given that both the segmentation accuracy and uncertainty were evaluated in the original Cartesian plane, the results in Figure 5(A) suggest that the adopted $d = 0.3$ is appropriate for the segmentation task. Although other $d$ values can be specified to obtain different uncertainty results, the projection calculations may limit the segmentation accuracy. Additionally, the number of projections $k$ can be arbitrarily selected. In this work, we set the projection center to be spaced at 8-pixel intervals, i.e., $k = 1024$. Based on Figure 5(C), this choice magnifies each sub-region in the original MRI by a factor $f_{SP}$ greater than 2, which is expected to provide sufficient projections for uncertainty quantification while balancing the computational cost. Generalizing the proposed method to other applications may result in a different optimal $k$, and the deep neural network architecture may require further optimization to support the increased computational demands at higher $k$.

## 5. Conclusion

In this work, we developed a segmentation uncertainty quantification method based on spherical projection for U-Net. Using a large database of multi-parametric MRI-based glioma segmentations, the developed technique achieved high segmentation accuracy and successfully highlighted missed tumor regions and potential segmentation errors. The presented methodology can be generalized to other medical image-related deep-learning applications for uncertainty evaluation.